\documentclass[conference, a4paper]{IEEEtran}

\IEEEoverridecommandlockouts          
\usepackage{booktabs}
\usepackage{graphicx}
\usepackage{multirow}
\usepackage{amsmath,amssymb,latexsym}
\usepackage{rotating}
\usepackage{lettrine}
\usepackage{textcomp}
\usepackage{units}
\usepackage[table]{xcolor}

\usepackage{hyperref}
\usepackage{tikz}

\usepackage{xcolor}

\usepackage{lipsum}% For 'Lorem Ipsum' dummy text
\usepackage[pscoord]{eso-pic}% The zero point of the coordinate systemis the lower left corner of the page (the default).
\newcommand{\placetextbox}[3]{% \placetextbox{<horizontal pos>}{<vertical pos>}{<stuff>}
  \setbox0=\hbox{#3}% Put <stuff> in a box
  \AddToShipoutPictureFG*{% Add <stuff> to current page foreground
    \put(\LenToUnit{#1\paperwidth},\LenToUnit{#2\paperheight}){\vtop{{\null}\makebox[0pt][c]{#3}}}%
  }%
}%

\title{Multi-Link Operation and Wireless Digital Twin to Support Enhanced Roaming in Next-Gen \mbox{Wi-Fi}
\thanks{This work has been partially funded by SoBigData.it ``SoBigData.it receives funding from European Union – NextGenerationEU – National Recovery and Resilience Plan (Piano Nazionale di Ripresa e Resilienza, PNRR) – Project: “SoBigData.it – Strengthening the Italian RI for Social Mining and Big Data Analytics” – Prot. IR0000013 – Avviso n. 3264 del 28/12/2021.'', and partially funded by the European Commission Horizon Europe SNS JU PREDICT-6G (GA 101095890) Project.}
}

\author{
    \IEEEauthorblockN{
    Stefano Scanzio\IEEEauthorrefmark{1},
    Matteo Rosani\IEEEauthorrefmark{1}\IEEEauthorrefmark{2},
    Gabriele Formis\IEEEauthorrefmark{1}\IEEEauthorrefmark{2},
    Dave Cavalcanti\IEEEauthorrefmark{3},\\
    Valerio Frascolla\IEEEauthorrefmark{4},
    Guido Marchetto\IEEEauthorrefmark{2},
    Gianluca Cena\IEEEauthorrefmark{1},
    }

    \IEEEauthorblockA{\IEEEauthorrefmark{1}National Research Council of Italy (CNR--IEIIT), Italy. \IEEEauthorrefmark{2}Politecnico di Torino, Italy.}    
    \IEEEauthorblockA{\IEEEauthorrefmark{3}Intel Labs, Intel Corporation, Hillsboro, OR, USA. \IEEEauthorrefmark{4}Intel Labs, Intel Deutschland, Neubiberg, Germany.}
    Email: name.surname@\{cnr.it, intel.com, polito.it\}
    }

\begin{document}
\placetextbox{0.5}{1}{This is the author's version of an article that has been published.}
\placetextbox{0.5}{0.985}{Changes were made to this version by the publisher prior to publication.}
\placetextbox{0.5}{0.97}{The final version of record is available at \href{https://doi.org/10.1109/WFCS60972.2024.10540931}{https://doi.org/10.1109/WFCS60972.2024.10540931}}%
\placetextbox{0.5}{0.05}{Copyright (c) 2024 IEEE. Personal use is permitted.}
\placetextbox{0.5}{0.035}{For any other purposes, permission must be obtained from the IEEE by emailing pubs-permissions@ieee.org.}%

\maketitle
\thispagestyle{empty}
\pagestyle{empty}

%%%%%%%%%%%%%%%%%%%%%%%%%%%%%%%%%%%%%%%%%%%%%%%%%%%%%%%%%%%%%%%%%%%%%%%%%%%%%%%%
\begin{abstract}
The next generation of \mbox{Wi-Fi} is meant to achieve ultra-high reliability for wireless communication.
Several approaches are available to this extent, some of which are being considered for inclusion in standards specifications, including coordination of access points to reduce interference.

In this paper, we propose a centralized architecture based on digital twins, called WiTwin, with the aim of supporting wireless stations in selecting the optimal association according to a set of parameters.
Unlike prior works, we assume that \mbox{Wi-Fi} 7 features like multi-link operation (MLO) are available.
Moreover, one of the main goals of this architecture is to preserve communication quality in the presence of mobility, by helping stations to perform reassociation at the right time and in the best way.
\end{abstract}

%%%%%%%%%%%%%%%%%%%%%%%%%%%%%%%%%%%%%%%%%%%%%%%%%%%%%%%%%%%%%%%%%%%%%%%%%%%%%%%%

\section{Introduction}
\label{sec:introduction}
Due to its high performance and low-cost, \mbox{Wi-Fi} is becoming one of the most pervasive communication technologies, which will enable new production paradigms in the factories of the future. 
The recent enhancements to IEEE 802.11 specifications make \mbox{Wi-Fi} suitable, in terms of reliability and communication latency, for many application contexts related to Industry 4.0 and the forthcoming Industry 5.0.
Some of these improvements, especially those aimed at enhancing network throughput and multi-link operation (MLO), have been included in the IEEE 802.11be amendment  (also known as \mbox{Wi-Fi} 7), while others are being defined by the IEEE 802.11bn Task Group for \mbox{Wi-Fi} 8 \cite{giordano2023wifi}.

When mobility is considered, e.g., for Automated Guided Vehicles (AGV) and Autonomous Mobile Robots (AMR),
which are increasingly used in modern factories, \mbox{Wi-Fi} does not perform as well as cellular technologies like 5G.
Roaming still constitutes a critical operation in \mbox{Wi-Fi} because, when a wireless station (STA) reassociates to a different access point (AP), it experiences a short interval (up to a few seconds) where communication quality is either unsatisfactory (excessive latency and jitters) or prevented altogether. 
These gaps are annoying for soft real-time applications, but may lead to serious consequences in firm/hard real-time ones (control loops), where the lack of communication is not tolerated.

The problem of channel selection in a dense deployment has been explored in \cite{10293832} using reinforcement learning techniques. 
In particular, a Parallel Transfer Reinforcement Learning (PTRL) algorithm has been presented to improve convergence of MLO-enabled networks in the event of a channel selection.
In \cite{place1} the channel allocation problem is explored in the Internet of Things context and a multi-armed-bandit-based channel allocation method is proposed to deliver higher frame delivery ratio (FDR) and faster response for AP channel selection.
Lastly, a channel selection method that relies on the network load is presented in \cite{place2}, which estimates the load of the network and uses this information to make intelligent decisions.
Another critical point is the handover procedure because the interruption of service is a challenging problem in time-sensitive scenarios \cite{DiGregorio2019}. In \cite{place3} a new technique is proposed to reduce the handover delay to a bounded value.

In this work, a digital twin (DT) of the wireless communication environment is defined that permits moving STAs with MLO capability to perform seamless roaming between APs. 
In particular, the proposed DT exploits the capability of multi-link devices (MLD) to have multiple 
links active at the same time.
This means that, during reassociation, links can be moved one at a time to the new AP,
so that at any time there is always at least one link that is operating properly and supports communication with adequate quality for time-aware applications.
The paper is structured as follows: Section~\ref{sec:system} sketches an overview of the system architecture, while its main operations and components are analyzed in Section~\ref{sec:operation}, just before the conclusive remarks are drawn in Section IV.

\section{System architecture}
\label{sec:system}

\begin{figure}[]
    \begin{center}    \includegraphics[width=1.0\columnwidth]{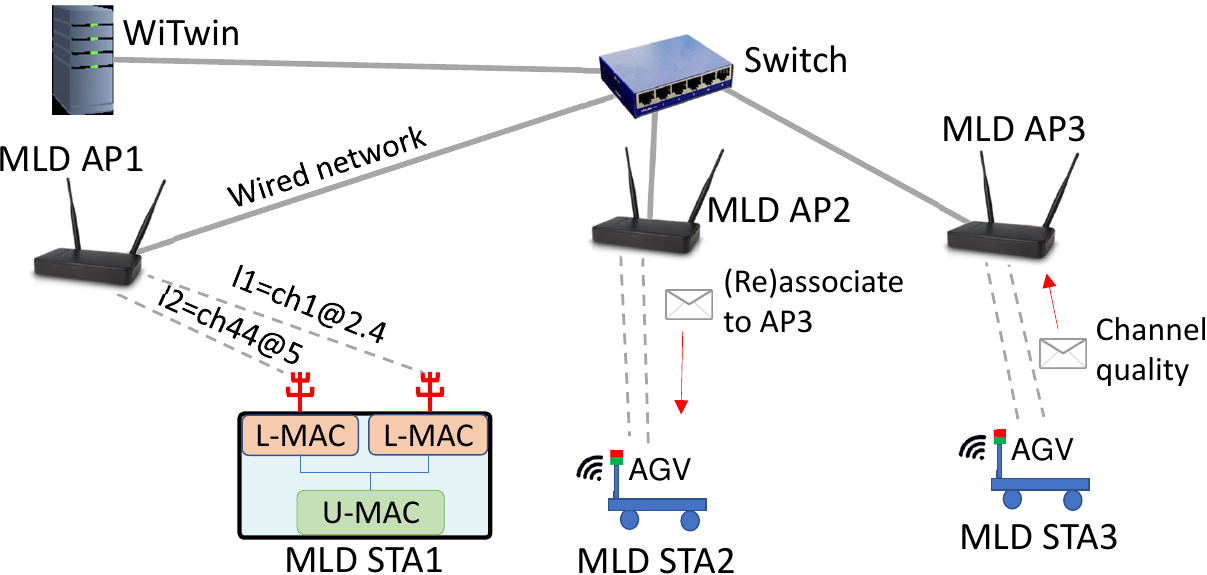}
    \end{center}
    \caption{Proposed architecture based on WiTwin.}
    \label{fig:arch}
\end{figure}

The architecture of Fig.~\ref{fig:arch}, which is based on a centralized approach, is being proposed to support reliable communication in presence of STA mobility, e.g., to interconnect fleets of AGVs and AMRs.
In the proposed architecture, APs are connected to a wired infrastructure (typically an Ethernet network), which permits them to communicate with a centralized entity named \textit{WiTwin}.
We assume that the wired network behaves ideally in terms of frame losses.
If fault-tolerance is demanded, solutions like parallel redundancy can be exploited.

Since we are witnessing the dawn of \mbox{Wi-Fi} 7, as specified by IEEE 802.11be \cite{9152055},
we assume that every AP is an MLD that supports MLO \cite{10149044,10012034}, i.e., it may have two or more active associations to different APs and transmit (quasi) simultaneously on different links tuned on distinct channels. 
Conversely, STAs could be either MLDs or legacy \mbox{Wi-Fi} devices.
In an MLD, the media access control (MAC) layer is split in two ``sub-layers'':
at the bottom L-MAC entities (two or more, each one related to a specific link) interface with the PHYs and manage aspects like channel access and retransmissions. 
On top the \mbox{U-MAC} (a single entity for every STA) coordinates all the local \mbox{L-MACs} and interface them with the data-link user. Every STA cyclically sends the perceived \textit{channel quality}, as well as its position, to the WiTwin. Starting from the information acquired from STAs, the WiTwin builds (and then keeps updated) a spatial model of the radio environment, described in terms of the communication quality for every channel.
Mobile STAs can exploit this model, named WiTwinModel, to optimize roaming between APs.
However, the same model can be also exploited by static STAs to balance the load among APs and improve the overall communication quality on air.

Roaming in legacy \mbox{Wi-Fi} is managed by every single STA separately, and decisions about reassociation rely only on locally observable quantities. Examples are receive signal strength indicator (RSSI) and FDR.
For the \textit{network-driven roaming} we propose, the WiTwin notifies the STAs the information about the new (best) AP to reassociate using specific messages.
In this case, all the available information, as described by the WiTwinModel (including the current position and mobility pattern of the STAs), can be exploited to make optimal decisions.
Among the metrics that can be used to select the new AP, we deem that latency could be one of the most appropriate in the context of mobility in industrial environments. 
In fact, for this class of applications, ensuring bounded transmission delays (even from a probabilistic point of view) is essential for the correct operation of machinery and equipment.
Reassociation can be triggered either directly by the WiTwin (\textit{proactive} behavior) or by the STA when the observed communication quality on the link suddenly falls below a given threshold (\textit{reactive} behavior).
In addition to network-driven roaming, legacy \textit{STA-driven roaming} is still needed to permit STAs to reassociate also when communication with the AP is prevented. 
In this case, the STA autonomously decides the best AP, possibly exploiting prior information that was made available by the WiTwin and cached in the STA, or by using machine learning (ML) algorithms aimed at predicting channel quality in the near future \cite{https://doi.org/10.1002/itl2.326}.

Concerning the above architecture, an important aspect is the \textit{reassociation procedure}. 
Specifically, reassociation of MLD STAs should be done one L-MAC at a time to keep latency bounded and preserve reliable communication with nodes connected to the wired network, e.g., a programmable logic controller (PLC) \cite{10144124}.

\section{Main operations and components}
\label{sec:operation}
Our proposal is characterized by four main aspects that are analyzed independently: 
feature acquisition, WiTwinModel, roaming, and MLD reassociation.

\subsection{Feature acquisition}
This aspect encompasses the acquisition and transfer of information from STAs, useful to build the WiTwinModel. 
The $i$-th sample
obtained on transmissions from the source STA $s$ to the destination AP $d$ (or vice-versa) on channel $c$, 
can be described as a tuple $x^{\langle s,d,c \rangle}_i=\langle \nu_1, \nu_2,...,\nu_{N} \rangle$, 
where $N$ is the dimension of the feature space, which depends on the measuring device's capabilities (i.e., on the ability of the node to measure certain physical quantities).
For example, an MLD STA with three L-MACs operating on channels in the $\unit[2.4]{GHz}$, $\unit[5]{GHz}$, and $\unit[6]{GHz}$ bands will acquire features related to any of its three channels, and
information on frame transmissions between STA 3 and AP 5 (unique identifiers are needed, e.g., MAC addresses) on channel 1 in the $\unit[2.4]{GHz}$ band will be denoted $x^{\langle 3, 5, ch1@2.4 \rangle}_i$.
For instance, in the sample $x^{\langle s,d,c \rangle}_i=\langle t, x, y, \nu_4,...,\nu_{N} \rangle$, 
$t$ is a timestamp taken using a common (network-wide) time base, $x$ and $y$ (plus, possibly, $z$) represent the position of the device, 
while $\nu_4,...,\nu_{N}$ are other features related to channel quality (e.g., FDR, 
RSSI, number of transmission attempts for a packet). 
For APs, the position needs not be refreshed often because they are static.
In the initial version of the architecture we suppose that nodes are localized by means of either the IEEE 802.11az protocol or other less precise techniques \cite{Roy2022}.

Delivering sample $x^{\langle s,d,c \rangle}_i$ from the STA to the WiTwin can be done 
by means of specifically-defined messages (that may include one or more samples, possibly compressed) or
by using messages defined in IEEE 802.11k (after the required adjustments).
Moreover, the AP may also infer some quantities by analyzing message exchanges on air (e.g., from the reception of beacon, data, and ACK frames). 
APs exploit the wired network to deliver information to the WiTwin.

\subsection{WiTwinModel}
The environment is modeled by WiTwin using the WiTwinModel. 
While exact details about it must still be finalized, it is based on a heatmap that, in its most straightforward implementation, is defined on a two-dimensional space (3D extensions can be easily obtained). 
For every point in the plane the WiTwinModel provides the estimated transmission channel quality $\hat h$
(or, possibly, its forecast in the immediate future) to each MLD AP in the area of interest. 
Depending on the roaming management algorithm (introduced in the next section), the estimate $\hat h$ can refer to different quantities. 
At present, RSSI is widely used in \mbox{Wi-Fi} STAs to decide whether and where to roam \cite{8001421}. 
In this work, we propose instead to use the estimated FDR, as it is directly related to the probability to succeed in sending a frame to destination, and consequently to latency.

The WiTwinModel is constantly (re)trained using the features acquired at runtime, and implements the function
\begin{equation}
    \hat h(x,y,id_{\mathrm{AP}},c),
\end{equation}
that, given the STA position ($x,y$ or $x,y,z$), a tag $id_{\mathrm{AP}}$ that unambiguously identifies an AP, and a specific channel $c$, 
returns an estimate of the link quality.
A possible extension of $\hat h$ is to add the future time $t_f$ at which to perform estimation, obtaining a slightly different prototype $\hat h(x,y,id_{\mathrm{AP}},c, t_f)$. 
Doing so permits to better handle mobility because the time when reassociation  is done
can be selected in advance based on spectrum dynamics and the mobility characteristics of the node (direction and speed), 
but it also makes the implementation of the WiTwinModel noticeably more complex.

We wish to point out that, in this architecture, MLD APs are not enabled to change at runtime their configuration (in terms of the used channels). 
In fact, this operation is quite complex in practice, as all nodes already associated with an AP would need to be migrated to other channels, which unavoidably impacts on the real-time characteristics of traffic.

\begin{figure}[t]
    \begin{center}
    \includegraphics[width=0.86\columnwidth]{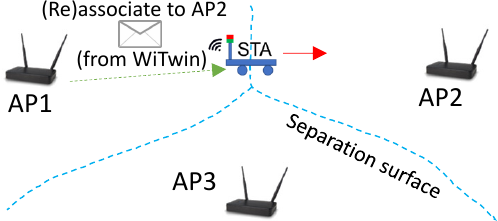}
    \end{center}
    \caption{Example of network-driven roaming.}
    \label{fig:roaming}
\end{figure}

As an example, for an MLD AP with three configured links (e.g., channel 5 in $\unit[2.4]{GHz}$ band, channel 44 in $\unit[5]{GHz}$ band, and channel 73 in $\unit[6]{GHz}$ band), the WiTwinModel may return for a given point in the plane (or in the space) a tuple of three elements that model the estimated channel quality related to a future position of a node.
In practice, this information can be obtained by invoking function $\hat h$ thrice, and results can be used by the WiTwin to compute and send information to the STA about the AP to which it must reassociate, plus additional information about the optimal roaming sequence. 
These aspects are discussed in the next section. 
Concerning the previous example, the tuple might be as follows
\begin{equation}
    \Bigl< \hat h(\cdot,ch5@2.4), \hat h(\cdot,ch44@5), \hat h(\cdot,ch73@6) \Bigr>,
\end{equation}
where the first three arguments of $\hat h$ ($x, y, id_{\mathrm{AP}}$) have been replaced with a dot because they are the same in all the calls.

Estimation carried out by $\hat h$ can be implemented using either statistical methods or ML. 
Both methods are trained on the set of feature $\{ x^{\langle s,d,c \rangle}_i \}$, where more weight should be clearly given to the most recent samples (the latter is typically handled automatically by ML algorithms). 
Concerning the use of ML for predicting the future channel quality, a good survey is presented in \cite{9786784}, including, e.g., the use of combinations of exponential moving averages \cite{10218083} and artificial neural networks \cite{https://doi.org/10.1002/itl2.326}.

\subsection{Roaming}

In this paper we only focus on network-driven roaming, even if the services provided by the WiTwin could greatly improve communication latency also for STA-driven roaming. 
In network-driven roaming, the WiTwin sends a specific message to the STA to enforce the reassociation procedure when the node enters (or is expected to) an area where another AP is available that is characterized by a higher FDR
(switching to it may potentially improve communication latency too). 
In this context, the WiTwin uses the WiTwinModel to determine in real-time when the channel quality of the new candidate AP overtakes the current AP.
The new AP $id_{\mathrm{AP'}}$ is selected as follows:
\begin{equation}
    id_{\mathrm{AP'}} = \arg \min_{a\in \mathcal{A}} \sum_{c \in \mathcal{C}^{a}} \Bigl( 1-\hat h(x, y, a, c) \Bigr),
\end{equation}
where $\mathcal{A}$ is the set of all the APs on which search is performed 
while $\mathcal{C}^{a}$ is the set of all the channels configured on AP $a$.
The message sent by the WiTwin to the STA to enforce reassociation is generated only when $id_{\mathrm{AP'}}$ differs from the current AP and improvements, in terms of the aggregate FDR (which considers all the links of the MLD AP) exceed a given threshold, to prevent oscillations and network instability.

This message is delivered to the STA either through the AP it is associated to or, possibly, via the beacons of neighboring APs operating on the same channel, and specifies the target AP it must/should reassociate and what is the migration order of the affiliated L-MACs.
With reference to the example of Fig.~\ref{fig:roaming}, when the WiTwin notices that the STA is going to enter into an area where a better communication quality (i.e., higher FDR) is available, AP1 sends a message instructing the STA to reassociate to AP2.

\subsection{MLD reassociation}
\begin{figure}[t]
    \begin{center}
    \includegraphics[width=1.0\columnwidth]{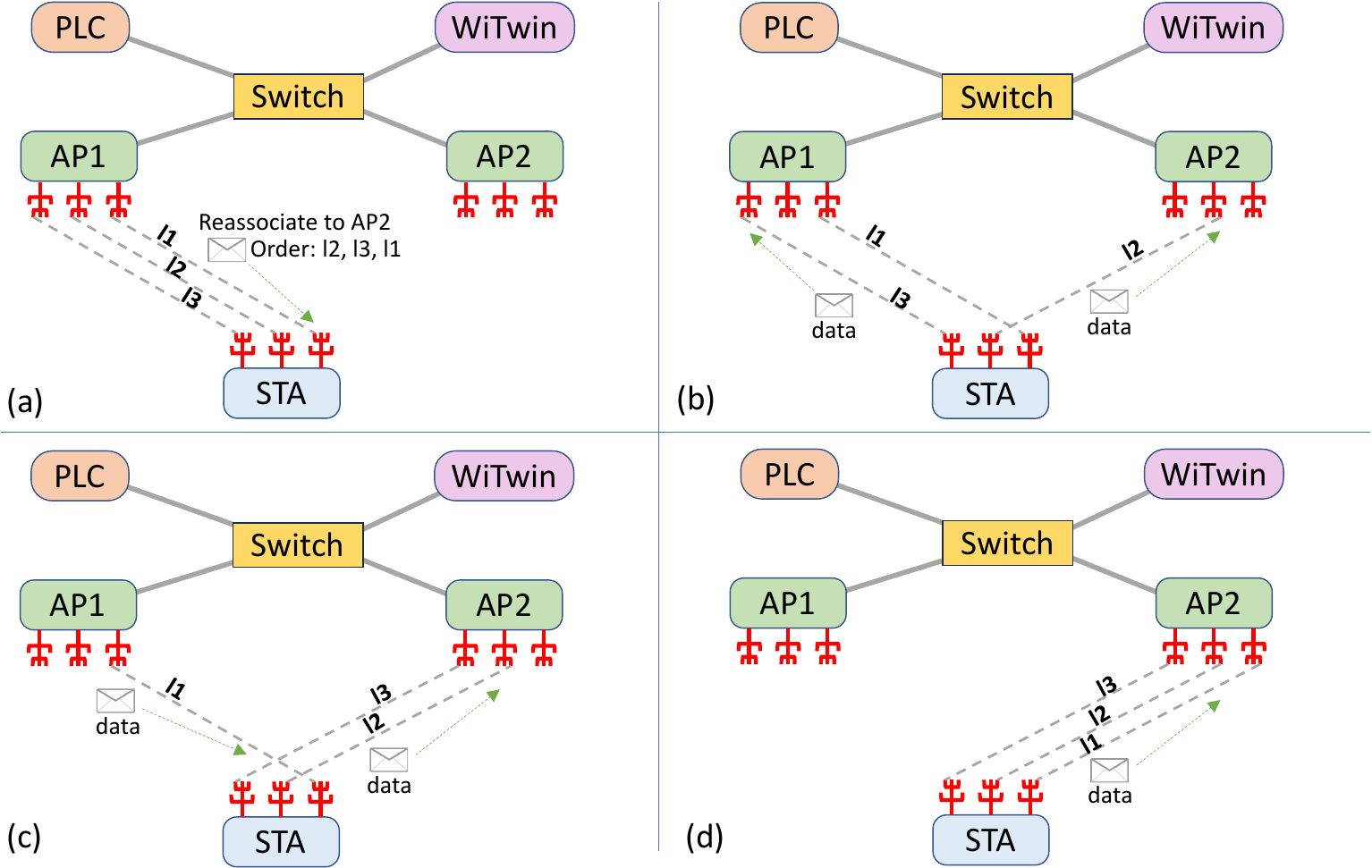}
    \end{center}
    \caption{Sequence of steps performed during MLD roaming.}
    \label{fig:MLD_roaming}
\end{figure}
Likely, managing the reassociation procedure of MLDs is one of the most interesting points discussed in this work.
To ensure bounded (and minimal) latency to the time-sensitive applications that involve a moving MLD STA, e.g., a PLC connected to the wired network controlling an AMR over the air, reassociation is assumed to always involve a single L-MAC at a time. 
During reassociation, for a certain time interval the MLD STA is associated with two distinct MLD APs (the older and the newer ones), with the U-MAC selecting frame-by-frame the best option.
In this case, packets could be possibly sent on paths involving both APs. 
By doing so, connection is virtually never interrupted.
The possibility of sending two copies of the same packet on multiple links at the same time permits to drastically reducing latency during roaming operations in time-critical \mbox{Wi-Fi} applications \cite{10144228}. 
However, it requires a more detailed analysis to determine which network entity is in charge of removing duplicates.

The message sent by WiTwin to the MLD STA that triggers reassociation also specifies the migration order. 
This sequence is selected to probabilistically maximize the combined link quality in terms of FDR.
The MLD roaming process is shown in Fig.~\ref{fig:MLD_roaming}. 
Let us assume (Fig.~\ref{fig:MLD_roaming}.a) that the moving STA is associated to AP1 with three links ($l_1, l_2, l_3$) to transmit/receive data to/from the PLC located in the wired network. 
When the WiTwin detects that, according to the law of motion, AP2 has (likely) become able to provide better communication quality than AP1, it sends a message to the STA that contains the indication of this new AP and the order in which links must be moved from the current to the new AP (e.g., $l_2$, $l_3$, $l_1$). 

At this point the STA starts moving link $l_2$ to the new AP2 (Fig.~\ref{fig:MLD_roaming}.b), but in the meanwhile, links $l_1$ and $l_3$ can be still used for communicating with the PLC. 
Then, it is the turn of $l_3$ (Fig.~\ref{fig:MLD_roaming}.c), and links $l_2$ a $l_1$ can be used for communication. 
In this case, part of the messages flow through AP1 and part through AP2. 
Finally, also the link $l_1$ is reassociated with AP2 (Fig.~\ref{fig:MLD_roaming}.d), and links $l_2$ and $l_3$ can be used in the meanwhile, after which the MLD roaming procedure is completed and the STA can use all the three links now associated with AP2.

\section{Conclusions}
\mbox{Wi-Fi} is currently the most popular wireless communication technology that enables wireless extensions to Ethernet network infrastructures.
One of the most effective approaches to increase its reliability is to coordinate APs operations to reduce the interference among them.
To this aim several proposals, which select associations between STAs and APs to maximize communication quality, have recently appeared.

In this paper we consider two additional aspects that are expected to gain attention in the close future.
First, we rely on \mbox{Wi-Fi} 7, whose specifications are going to be published soon and for which commercial equipment has started to be available.
Second, we focus on mobility of STAs, and strive to optimize roaming (i.e., the handover procedure) by improving the way reassociation is carried out.

In particular, we exploit a DT for the radio environment that finds: a) the best time to perform reassociation, before communication quality starts to worsen sensibly; b) the best option for the next AP to which reassociate; and, c) in which order the different links of an MLD must be moved to the new AP.
To this purpose we are working on the definition of a suitable architecture, which we present here in its draft form.

\bibliographystyle{IEEEtran}
\bibliography{bibliography}

\end{document}